\documentclass[12pt]{article}
\usepackage{amsmath,epsfig}
\makeatletter
\def\Journal#1#2#3#4{{#1} {\bf #2}, #3 (#4)}

\def\NPB{{\em Nucl. Phys.} B}
\def\PLB{{\em Phys. Lett.}  B}
\def\PRL{\em Phys. Rev. Lett.}
\def\PRD{{\em Phys. Rev.} D}
\def\ZPC{{\em Z. Phys.} C}
\def\ZPA{{\em Z. Phys.} A}
\def\NPA{{\em Nucl. Phys.} A}

\def\ra{\rightarrow}

\def\al{\alpha}
\def\be{\begin{equation}}
\def\ee{\end{equation}}
\def\bea{\begin{eqnarray}}
\def\eea{\end{eqnarray}}
\def\rd{{\rm d}}
\def\mev{\,{\rm MeV}}
\def\gev{\,{\rm GeV}}

\newcommand{\das}{distribution amplitudes}

\begin{document}
\sloppy
\pagestyle{empty}

\begin{flushright}
WU B 98-22 \\
hep-ph/9807471\\[6mm]
\end{flushright}

\begin{center}
{\Large\bf The Diquark Model for Exclusive Reactions}\\[10mm]

{\Large P.\ Kroll\footnote
{Invited talk presented at the Workshop on N* physics and non-perturbative QCD, 
Trento (1998)}
\footnote{Supported in part by the TMR network ERB 4061 Pl 95 0115}
}\\[3mm]
{\it Fachbereich Physik, Universit\"at Wuppertal}\\[1mm]
{\it Gau\ss strasse 20, D-42097 Wuppertal, Germany}\\[1mm]%
{ E-mail: kroll@theorie.physik.uni-wuppertal.de}\\[10mm]
\end{center}
\begin{abstract}
The present status of the diquark model for exclusive 
reactions at moderately large momentum transfer is reviewed. That 
model is a variant of the Brodsky-Lepage approach in which diquarks are
considered as quasi-elementary constituents of baryons. Recent
applications of the diquark model, relevant to high energy physics with
electromagnetic probes, are discussed.
\end{abstract}

\section{Introduction}
Exclusive processes at large momentum transfer are described in terms
of hard scatterings among quarks and gluons \cite{lep:80}. In this
so-called hard scattering approach (HSA) a hadronic amplitude
is represented by a convolution of process independent
\das{} (DA) with hard scattering amplitudes to be calculated within
perturbative QCD. The DAs specify the distribution
of the longitudinal momentum fractions the constituents carry. They
represent Fock state wave functions integrated over transverse
momenta. The convolution manifestly factorizes long (DAs) and short
distance physics (hard scattering). It however turned out that most
processes are not dominated by the perturbative contribution at
experimentally accessible values of momentum
transfer. Non-perturbative dynamics still plays a crucial role in that
kinematical region and, hence, the HSA although likely the correct
asymptotic picture for exclusive reactions, needs modifications.

In a series of papers \cite{ans:87}-\cite{kro:96a} 
such a modification has been proposed in which baryons are
viewed as being composed of quarks and diquarks. The latter are treated as 
quasi-elementary constituents which partly survive medium hard
collisions. Diquarks are an effective description of correlations in 
the wave functions and constitute a particular model for
non-perturbative effects. The diquark model may be viewed as a variant
of the HSA appropriate for moderately large momentum transfer and it 
is designed in such a way that it evolves into the standard pure quark
HSA asymptotically. In so far the standard HSA and the diquark model 
do not oppose each other, they are not alternatives but rather complements.
The existence of diquarks is a hypothesis. However, from experimental
and theoretical approaches there have been many indications suggesting
the presence of diquarks. For instance, they were introduced in baryon
spectroscopy, in nuclear physics, in astrophysics, in jet fragmentation 
and in weak interactions to explain the famous $\Delta I=1/2$ rule. 
Diquarks also provide a natural explanation of the equal slopes of 
meson and baryon Regge trajectories. For more details and for 
references, see \cite{kro:87}. It is important to note that QCD
provides some attraction between two quarks in a colour $\{\bar{3}\}$ 
state at short distances as is to be seen from the static reduction of the
one-gluon exchange term. 

Even more important for our aim, diquarks have also been found to play a role in
inclusive hard scattering reactions. The most obvious place to signal
their presence is deep inelastic lepton-nucleon scattering. Indeed
the higher twist contributions, convincingly observed by the NMC 
\cite{vir:91}, can be modelled as lepton-diquark elastic scattering. 
Baryon production in inclusive $pp$ collisions also reveals the need 
for diquarks scattered elastically in the hard interaction
\cite{szc:92}. For instance, kinematical dependences or the excess of
the proton yield over the antiproton yield find simple explanations
in the diquark model. No other explanation of these phenomena is
known as yet.
\section{The Diquark Model}
As in the standard HSA a helicity amplitude
for the reaction $AB\ra CD$ is expressed as a convolution of DAs and
hard scattering amplitudes ($s$, $-t$, $-u$\,\, $\gg m_i^2$) 
\begin{eqnarray}
\label{helamp}
M (s,t)\,=\,\int \rd x_C \rd x_D \rd x_A \rd x_B  
\Phi^*_C(x_C) \Phi^*_D(x_D) T_H(x_i,s,t) \Phi_A(x_A) \Phi_B(x_B) 
\end{eqnarray}
where helicity labels are omitted for convenience. Implicitly
it is assumed in (\ref{helamp}) that the valence Fock states consist
of only two constituents, a quark and a diquark (antiquark) in the
case of baryons (mesons). In so far the specification of the quark
momentum fraction $x_i$ suffices; the diquark (antiquark) carries the
momentum fraction $1-x_i$. If an external particle is point-like,
e.g. a photon, the corresponding DA is to be replaced by $\delta (1-x_i)$.
As in the standard HSA contributions from higher Fock states are
neglected. This is justified by the fact that that such contributions 
are suppressed by powers of $\al_s/t$ as compared to that from the 
valence Fock state (if only S-wave hadrons are involved).

In the diquark model spin $0$ ($S$) and spin $1$ ($V$) colour
antitriplet diquarks are considered. Within flavour SU(3) the $S$ 
diquark forms an antitriplet, the $V$ diquark an sextet. 
Assuming zero relative orbital angular momentum between quark and
diquark and taking advantage of the collinear approximation, the
valence Fock state of a ground state octet baryon $B$ with helicity
$\lambda$ and momentum $p$ can be written in a covariant fashion 
(omitting colour indices)
\begin{equation}
\label{pwf}
|B;p,\lambda\rangle  = f_S\,\Phi_S^B(x)\,B_S\, u(p,\lambda) 
             + f_V\, \Phi_V^B(x)\, B_V
              (\gamma^{\alpha}+p^{\alpha}/m_B)\gamma_5 \,u(p,\lambda)/\sqrt{3}
\end{equation}
where $u$ is the baryon's spinor. The two terms in (\ref{pwf})
represent configurations consisting of a quark and either a scalar or a 
vector diquark, respectively. The couplings of the diquarks 
with the quarks in a baryon lead to flavour functions which e.g.\ for
the proton read
\vspace*{-0.1cm}
\begin{equation}
\label{fwf}
B_S=u\, S_{[u,d]}\hspace{2cm} 
B_V= [ u V_{\{u,d\}} -\sqrt{2} d\, V_{\{u,u\}}]/\sqrt{3}\, .
\end{equation}
The DAs $\Phi^B_{S(V)}$ are conventionally
normalized as $\int \rd x \Phi = 1$. The constants $f_{S}$ and $f_V$
play the role of the configuration space wave functions at the origin. \\
The DAs containing the complicated non-perturbative bound state physics, 
cannot be calculated from QCD at present. It is still
necessary to parameterize the DAs and to fit the eventual free
parameters to experimental data. Hence, both the models, the standard
HSA as well as the diquark model, only get a predictive power when a
number of reactions involving the same hadrons is investigated. In
the diquark model the following DAs have been proven to work
satisfactorily well in many applications \cite{jak:93}-\cite{kro:96a}:
\begin{eqnarray}
\label{a10}
\Phi^B_S(x)&\hspace{-0.3cm}=&\hspace{-0.3cm}N^B_S x (1-x)^3 
                                 \exp{\left[-b^2 (m^2_q/x + m^2_S/(1-x))\right]}\\
\Phi^B_V(x)&\hspace{-0.3cm}=&\hspace{-0.3cm}N^B_V x (1-x)^3 (1+5.8\,x - 12.5\,x^2)
\exp{\left[-b^2 (m^2_q/x + m^2_V/(1-x))\right]}. \nonumber 
\end{eqnarray}
The constants $N^B_S$ and $N^B_V$ are fixed through the normalization  
(e.g.\ for the proton  $N_S^p = 25.97$, $N_V^p = 22.92$). The DAs
exhibit a mild flavour dependence via the exponential whose other purpose 
is to guarantee a strong suppression of the end-point regions. 
The parameters appearing in the exponentials are not considered as
free parameters since the final results (form factors, amplitudes)
depend on their actual values only mildly. The following values for
the parameters are chosen: $b = 0.498\gev^{-1}$, $m_u=m_d=350 \mev$,
$m_S=m_V=580 \mev$. It is to be stressed that the quark and diquark
masses only appear in the DAs (\ref{a10}); in the hard scattering 
kinematics they are neglected.

The hard scattering amplitudes $T_H$, determined by short-distance
physics, are calculated from a set of Feyman graphs relevant to a
given process. Diquark-gluon and diquark-photon vertices appear in
these graphs which, following standard prescriptions, are defined as 
\begin{eqnarray}
\label{vert}
\mbox{S$\,$g$\,$S}:&& \phantom{-}i\,g_s t^{a}\,(p_1+p_2)_{\mu} \nonumber\\
\mbox{VgV}:&& -i\,g_{s}t^{a}\, 
\Big\{
 g_{\alpha\beta}(p_1+p_2)_{\mu}
- g_{\beta\mu}\left[(1+\kappa)\,p_2-\kappa\, p_1\right]_{\alpha} \nonumber\\
&&\quad\quad\quad - g_{\mu\alpha} \left[(1+\kappa)\,p_1-\kappa\, p_2\right]_{\beta} 
\Big\} 
\end{eqnarray}
where $g_s=\sqrt{4\pi\alpha_s}$ is the QCD coupling constant.
$\kappa$ is the anomalous magnetic moment of the vector diquark and 
$t^a=\lambda^a/2$ the Gell-Mann colour matrix. For the coupling of 
photons to diquarks one has to replace $g_s t^a$ by $-\sqrt{4\pi\alpha} e_D$ 
where $\alpha$ is the fine structure constant and $e_D$ is the electrical 
charge of the diquark in units of the elementary charge. The couplings 
$DgD$ are supplemented by appropriate contact terms required by 
gauge invariance.\\
The composite nature of the diquarks is taken into 
account by phenomenological vertex functions. Advice for the parameterization 
of the 3-point functions (diquark form factors) is 
obtained from the requirement that asymptotically the diquark 
model evolves into the standard HSA. Interpolating smoothly  between 
the required asymptotic behaviour and the conventional value of 1 at $Q^{2}=0$, 
the diquark form factors are actually parametrized as
\begin{eqnarray}
\label{fs3}
\vspace*{-0.5cm}
F_{S}^{(3)}(Q^{2})=\frac{Q_{S}^{2}}{Q_{S}^{2}+Q^{2}}\,,\qquad
F_{V}^{(3)}(Q^{2})=\left(\frac{Q_{V}^{2}}{Q_{V}^{2}+Q^{2}}\right)^{2}\,.
\end{eqnarray}
The asymptotic behaviour of the diquark form factors and the connection to 
the hard scattering model is discussed in more detail in 
Ref.\ \cite{kro:87,kro:91}. In accordance with the required asymptotic
behaviour the $n$-point functions for $n\geq 4$ are parametrized as
\begin{eqnarray}
\label{fsn}
\vspace*{-0.8cm}
F_{S}^{(n)}(Q^{2})=a_{S}F_{S}^{(3)}(Q^{2})\,,\qquad
F_{V}^{(n)}(Q^{2})=
\left(a_{V}\frac{Q_{V}^{2}}{Q_{V}^{2}+Q^{2}}\right)^{n-3}F_{V}^{(3)}(Q^{2}).
\end{eqnarray}
The constants $a_{S,V}$ are strength parameters. Indeed, since the diquarks in 
intermediate states are rather far off-shell one has to consider 
the possibility of diquark excitation and break-up. Both these possibilities 
would likely lead to inelastic reactions. Therefore, we have not to consider 
these possibilities explicitly in our approach but excitation and break-up 
lead to a certain amount of absorption which is taken into account by the 
strength parameters. Admittedly, that recipe is a rather crude
approximation for $n\geq 4$. Since in most cases the contributions from the
n-point functions for $n\geq 4$ only provide small corrections to
the final results that recipe is sufficiently accurate.

The diquark hypothesis
has striking consequences. It reduces the effective number of
constituents inside baryons and, hence, alters the power laws. 
In elastic baryon-baryon scattering, for instance, the usual power
$s^{-10}$ becomes $s^{-6} F(s)$ where $F$ represents the net effect of
diquark form factors. Asymptotically $F$ provides the missing four
powers of $s$. In the kinematical region in which the diquark model 
can be applied ($-t$, $-u \geq 4 \gev^2$), the diquark form factors 
are already active, i.e.\ they supply a substantial $s$ dependence 
and, hence, the effective power of $s$ lies somewhere between 6 and 10.
The hadronic helicity is not conserved in the diquark model at finite
momentum transfer since vector diquarks can flip their helicities when
interacting with gluons. Thus, in contrast to the standard HSA,
spin-flip dependent quantities like the Pauli form factor of the 
nucleon can be calculated.
\section{Electromagnetic Nucleon Form Factors}
This is the simplest
application of the diquark model and the most obvious place to fix the
various parameters of the model. The Dirac and Pauli form factors of
the nucleon are evaluated from the convolution formula (\ref{helamp}) 
with the DAs (\ref{a10}) and the parameters are determined from a best
fit to the data in the space-like region. The following set of parameters
\begin{equation}
\label{c1}
\begin{array}{cccc}
 f_S= 73.85\,\mbox{MeV},& Q_S^2=3.22 \,\mbox{GeV}^2, & a_S=0.15, &  \\
 f_V=127.7\,\mbox{MeV},& Q^2_V=1.50\,\mbox{GeV}^2, & a_V=0.05,&\kappa=1.39\,;
\end{array}
\end{equation}
provides a good fit of the data \cite{jak:93}. $\alpha_s$ is evaluated with 
$\Lambda_{QCD}=200\mev$ and restricted to be smaller than $0.5$. The 
parameters $Q_S$ and $Q_V$, controlling the size of the diquarks,
are in agreement with the higher-twist effects observed in the structure 
functions of deep inelastic lepton-hadron scattering \cite{vir:91} if these 
effects are modelled as lepton-diquark elastic scattering. The Dirac
and the Pauli form factors of the proton are very well reproduced. 
The predictions 
for the two neutron form factors are also in agreement with the data. 
However, more accurate neutron data are needed in the $Q^2$ region of 
interest in order to determine the model parameters better. 
The nucleon's axial form factor \cite{jak:93} and its electromagnetic
form factors in the time-like regions \cite{kro:93a} have also been 
evaluated. Both the results compare well with data. Even 
electroexcitation of nucleon resonances has been investigated 
\cite{kro:92,bol:94}. In the case of the $N\Delta$ form factor the
model results agree very well with the data presented in \cite{stu96}
while the model seems to provide to large values for the Coulomb form
factor \cite{bur95}.
\section{Real Compton Scattering (RCS)}
\begin{figure}[t]
\setlength{\unitlength}{1mm}
\begin{picture}(160,60)
 \put(30,0)
    {\psfig{figure=cler_fig3.ps,%
        bbllx=50pt,bblly=90pt,bburx=510pt,bbury=670pt,%
        height=5cm,clip=}}
 \put(85,3)
    {\psfig{figure=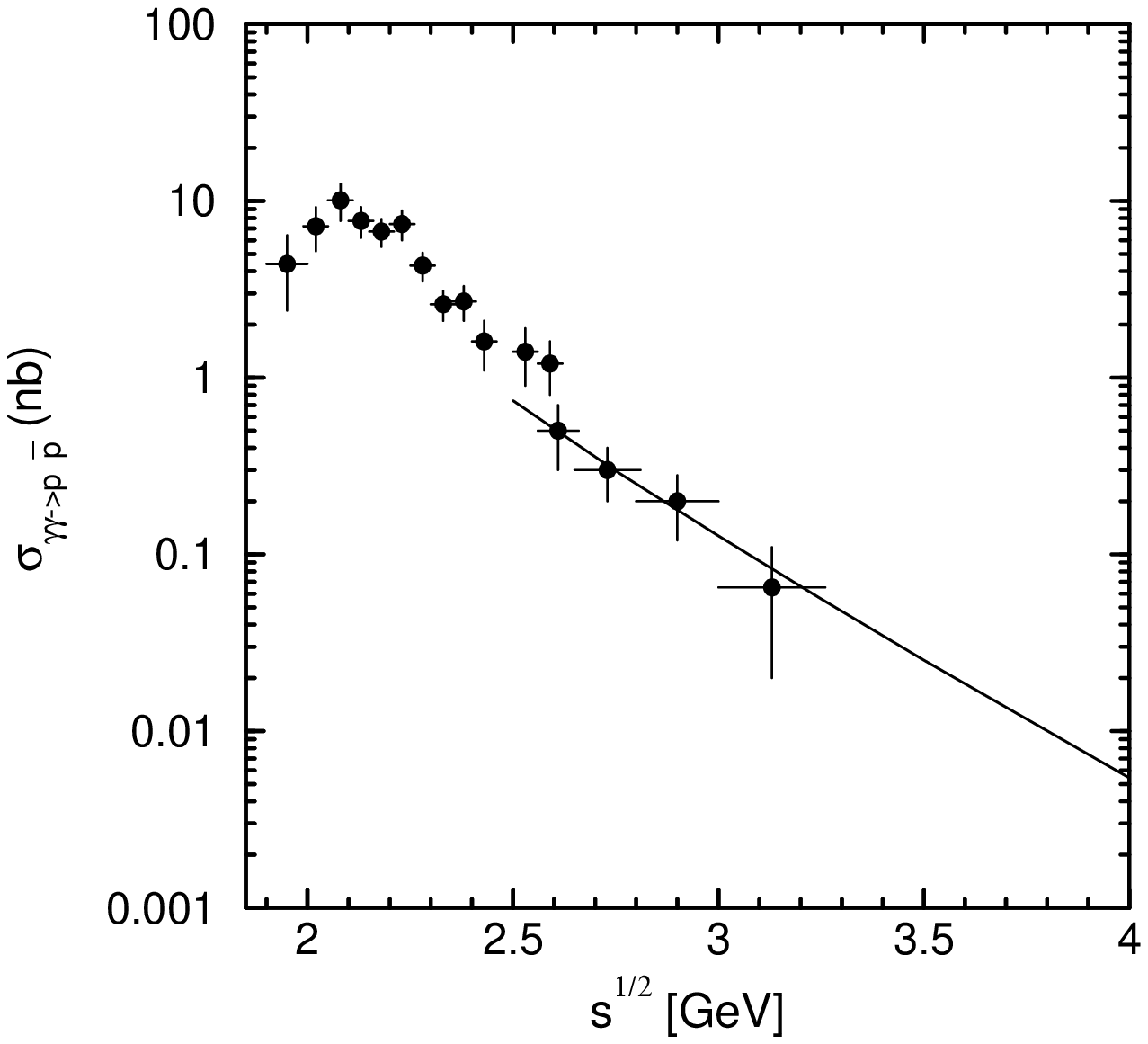,%
        bbllx=70pt,bblly=220pt,bburx=470pt,bbury=573pt,%
        height=4cm,clip=}}
\end{picture}
\caption[]{(left) The scaled cross section for RCS off protons  
vs.\ $\cos\theta$ for three different photon energies. 
The experimental data are taken from \cite{shu:79}.}
\label{frcc}
\vspace*{-0.3cm} 
\caption[]{(right) The integrated $\gamma\gamma\ra p\bar{p}$ cross section
  ($\mid\cos{\theta}\mid\ge 0.6$). The solid line represents the diquark
  model prediction \cite{kro:93a}. Data are taken from CLEO \cite{cleo}.} 
\label{time}
\vspace*{-0.5cm} 
\end{figure}

$\gamma p\ra\gamma p$ is 
the next reaction to which the diquark model is applied to. Since the 
only hadrons involved are again protons RCS can be predicted
in the diquark model without any adjustable parameter. 
The results of the diquark model for RCS are shown in Fig.\ \ref{frcc}{}
for three different photon energies \cite{kro:91,kro:96a}. 
Note that in the very forward and backward regions the transverse
momentum of the outgoing photon is small and, hence, the diquark
model which is based on perturbative QCD, is not applicable. Despite
the rather small energies at which data \cite{shu:79} are available, 
the diquark model is seen to work rather well. The predicted
cross section does not strictly scale with $s^{-6}$. The results 
obtained within the standard HSA are of similar quality
\cite{niz:91}. A purely soft, overlap-like contribution can also
explain these data \cite{rad:97}.
The diquark model also predicts interesting photon 
asymmetries and spin correlation parameters (see the discussion in 
\cite{kro:91}). Even a polarization of the proton, of the order of 
$10\%$, is obtained \cite{kro:91}. This comes about as a consequence 
of helicity flips generated by vector diquarks and of perturbative
phases produced by propagator poles appearing within the domains of
the momentum fraction integrations. The appearance of 
phases to leading order of $\al_s$ is a non-trivial 
prediction of perturbative QCD \cite{far:89}; it is characteristic 
of the HSA and is not a consequence of the diquark hypothesis.

Two-photon annihilation into $p\bar{p}$ pairs
is related to RCS by crossing. The only difference is that
now the diquark form factors are needed in the time-like region. 
The continuation of the diquark form factors from the space-like to
the time-like region is described in \cite{kro:93a}.
The diquark model predictions for the
integrated $\gamma\gamma\ra p\bar{p}$ cross section is compared to the
CLEO data \cite{cleo} in Fig.\ \ref{time}. At large energies the agreement
between predictions and experiment is good. The predictions for the 
angular distributions are in agreement with the CLEO data too. 
The diquark model predictions are also in agreement with the recent
VENUS data \cite{ven:97}.
\section{Virtual Compton scattering (VCS)}
This process is accessible
through $ep \ra ep \gamma$. An interesting element in that reaction 
is that, besides VCS, there is also a contribution from the 
Bethe-Heitler (BH) process where the final state photon is emitted 
from the electron. Electroproduction of photons offers many 
possibilities to test details of the dynamics: One may measure the 
$s$, $t$ and $Q^2$ dependence as well as that on the angle $\phi$ 
between the hadronic and leptonic scattering planes. This allows to
isolate cross sections for longitudinal and transverse virtual
photons. One may also use polarized beams and targets and last but not
least one may measure the interference between the BH and the VC 
contributions. The interference is sensitive to phase differences.
 
At $s$, $-t$ and $-u \gg m_p^2$ (or small $|\cos{\theta}|$ where
$\theta$ is the scattering angle of the outgoing photon in the
photon-proton center of mass frame) the diquark model can also be 
applied to VCS \cite{kro:96a}. Again there is no free parameter in 
that calculation. The model can safely be applied for 
$s\ge 10\gev^2$ and $|\cos{\theta}|\le 0.6$. For the future
CEBAF beam energy of $6\gev$ the model is at its limits of applicability. 
However, since the diquark model predictions for real Compton scattering 
agree rather well with the data even at $s\ge 5\gev^2$ (see Fig.\ \ref{frcc})
one may expect similarly good agreement for VCS. Predictions for the VCS 
cross section are given in \cite{kro:96a}. 

\begin{figure}[t]
\[
    \psfig{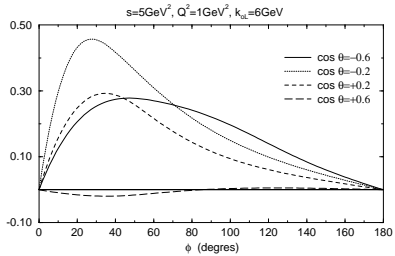}
\]
\vspace*{-1.0cm}
\caption[dummy5] {Diquark model predictions for the electron asymmetry
                  in $ep\to ep\gamma$ \cite{kro:96a}.}
\label{fig:asym}
\vspace*{-0.5cm}
\end{figure}

Of interest is also the electron asymmetry in 
$e p \to e p \gamma$:
\begin{equation}
  \label{asy}
A_L\,=\,\frac{\sigma(+) -\sigma(-)}{\sigma(+)+\sigma(+)}
\end{equation}
where $\pm$ indicates the helicity of the incoming electron. $A_L$
measures the imaginary part of the longitudinal -- transverse 
interference. 
According to the model, $A_L$ is large in the region of strong BH 
contamination (see Fig.\ \ref{fig:asym}). In that region, $A_L$ 
measures the relative phase between the BH amplitudes
and the VCS ones. The magnitude of the effect shown in
Fig.\ \ref{fig:asym} is sensitive to details of the model and,
therefore, should not be taken literally. Despite of this our results 
may be taken as an example of what may happen. The measurement of 
$A_L$, e.g.\ at CEBAF, will elucidate the underlying 
dynamics of VCS strikingly. 
\section{Summary and outlook}
The diquark model which represents a variant of the HSA,
combines perturbative QCD with non-perturbative elements. The diquarks
represent quark-quark correlations in baryon wave functions which are
modelled as quasi-elementary constituents. This model has been applied
to many photon induced exclusive processes at moderarely large
momentum transfer (typically $\simeq 4 \gev^2$). From the analysis of
the nucleon form factors the parameters specifying the diquark and the
DAs, are fixed. Compton scattering and two-photon annihilations
of $p\bar{p}$ can then be predicted. The comparison with existing
data reveals that the diquark model works quite well and in fact much
better then the pure quark HSA. 

Predictions for the VCS cross section and for the $ep\to ep\gamma$ 
cross section have also been made for kinematical situations
accessible at the upgraded CEBAF and perhaps at future high energy 
accelerators like ELFE@HERA. According to the diquark model the BH 
contamination of the photon electroproduction becomes sizeable for 
small azimuthal angles. The BH contribution also offers
the interesting possibility of measuring the relative phases
between the VC and the BH amplitudes. The electron
asymmetry $A_L$ is particularly sensitive to relative phases.
In contrast to the standard HSA the diquark model allows to
calculate helicity flip amplitudes, the helicity sum rule 
does not hold at finite $Q^2$. One example of an observable controlled
by helicity flip contributions is the Pauli form factor of the
proton. Also in this case the diquark model accounts for the data.


\end{document}